\documentclass[aip,rsi,reprint,graphicx]{revtex4-1} 

\draft 

\usepackage{graphicx}
\usepackage{color}
\usepackage{epstopdf} 
\usepackage{comment} 
\usepackage{textcomp}
\usepackage{wasysym}

\begin{document}


\title{Photoelectron Emission from Metal Surfaces Induced by Radiation Emitted by a 14~GHz Electron Cyclotron Resonance Ion Source} 
\thanks{Contributed paper, published as part of the Proceedings of the 16th International Conference on Ion Sources, New York City, NY, August, 2015.\\}



\author{Janne Laulainen}
\email[]{janne.p.laulainen@student.jyu.fi}
\author{Taneli Kalvas}
\author{Hannu Koivisto}
\author{Jani Komppula}
\author{Risto Kronholm}
\author{Olli Tarvainen}
\affiliation{University of Jyvaskyla, Department of Physics, P.O. Box 35, FI-40014 University of Jyvaskyla, Finland.}


\date{\today}

\begin{abstract}
Photoelectron emission measurements have been performed using a room-temperature $14$~GHz ECR ion source. It is shown that the photoelectron emission from Al, Cu, and stainless steel (SAE 304) surfaces, which are common plasma chamber materials, is predominantly caused by radiation emitted from plasma with energies between $8$~eV and $1$~keV. Characteristic X-ray emission and bremsstrahlung from plasma have a negligible contribution to the photoelectron emission. It is estimated from the measured data that the maximum conceivable photoelectron flux from plasma chamber walls is on the order of $10$~\% of the estimated total electron losses from the plasma.

\end{abstract}


\maketitle 

\section{Introduction}
Wall conditions can have a significant effect on the performance of electron cyclotron resonance ion sources (ECRIS) intended for high charge state production~\cite{Geller1996, Drentje2003}. Electron donors can improve the electron density of the plasma and low energy electrons emitted from the wall can have a significant effect on the plasma potential. The beneficial effect of wall coating was discovered when currents of highly charged oxygen ion beams increased by a factor of 2 after the plasma chamber had been covered by a silicon oxide layer~\cite{Lyneis1987}. Carbon contamination on the plasma chamber walls has been seen to affect the plasma potential and the oxygen ion beam current significantly~\cite{Tarvainen2005}. It is commonly believed that wall coating changes the plasma properties due to alteration of the secondary electron emission yield. Another possible process that can affect the electron emission from plasma chamber walls is photoelectric effect.

ECR plasmas emit radiation on the entire electromagnetic spectrum from radio waves to X-rays. Photons impinging a metal surface on the plasma chamber wall can lead to emission of photoelectrons (PE) when the energy of the photons exceeds the surface work function of the wall material. Photons carrying enough energy to cause PE emission from common metals are in the ultraviolet (UV) and X-ray parts of the spectrum. UV and X-ray photons are created in ECR plasmas due to electronic transitions from excited states to lower states of ions and neutrals and due to plasma bremsstrahlung. In ionizing plasmas the volumetric photon emission rate is directly proportional to the density of free electrons, ions and neutral particles and to the rate coefficient of the excitation reactions which depends on the electron energy distribution. A considerable part of the radiation emitted by ECRISs comes also from the walls due to wall bremsstrahlung and recombination of highly charged ions. As highly charged ions become neutralized at the walls by removal of electrons from the surface, the captured electrons mostly go into highly excited states creating a ``hollow ion'' with empty core states, which then leads to photon emission~\cite{Parks1995}.

Energy and material dependent quantum efficiency (QE) describes the average number of electrons emitted per incident photon. Vacuum-ultraviolet (VUV) is essential for PE emission since common metals have high QE at the VUV-part of the emission spectrum~\cite{Cairns1966}. Previous work with light ion sources suggest that PE emission could contribute to properties of hydrogen ion source plasmas~\cite{Laulainen2015}. ECR plasmas emit less intense VUV-radiation compared to light ion sources, since typical ECRISs operate at lower neutral gas pressure. However, ECR plasmas emit characteristic X-rays and  bremsstrahlung, and common metals have high QE at the X-ray parts of the emission spectrum as well~\cite{Henneken2000}.

\begin{figure*}[tb]
\includegraphics[width=1.0\textwidth]{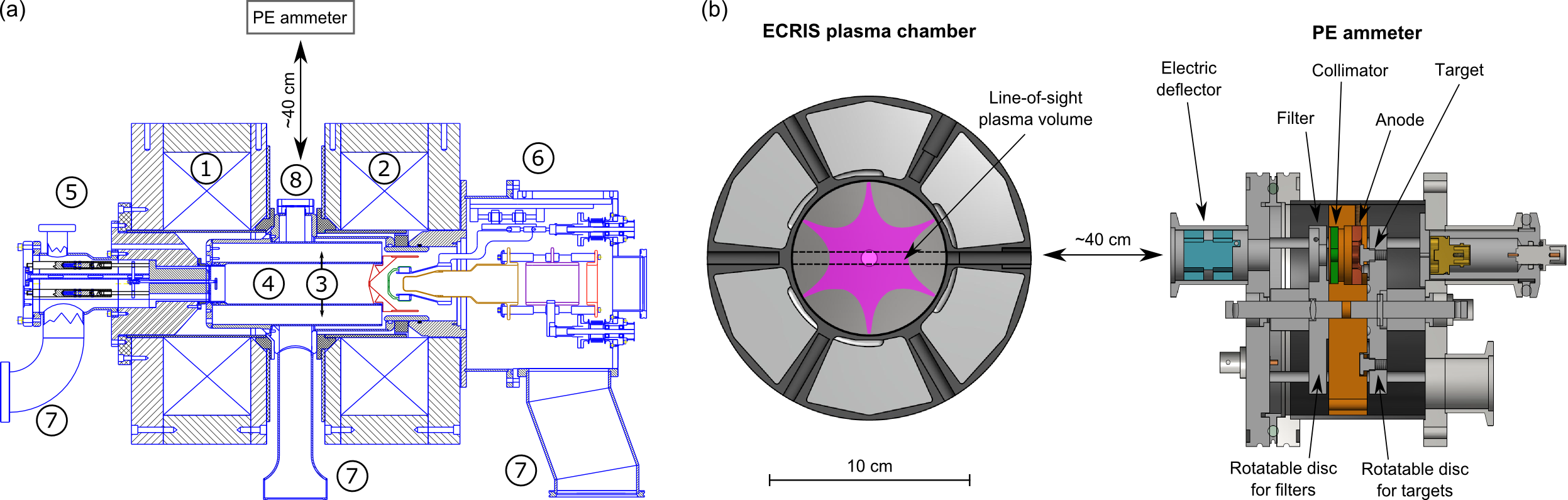}
\caption{(a) The JYFL 14 GHz ECRIS. (1) Injection and (2) extraction coils, (3) NdFeB sextupole, (4) plasma chamber ($\ell = 270$~mm, $\diameter = 76$~mm), (5) iron plug, biased disc, gas and microwave injection, (6) extraction system, (7) pumping, (8) vacuum gauge and PE ammeter. (b) ECRIS plasma chamber and the PE current measurement device.\label{fig:setup}}
\end{figure*}


\section{Measurement setup}
The measurement setup is presented in figure~\ref{fig:setup}. The experimental data presented in this article were taken with the room-temperature AECR-U type JYFL 14~GHz ECRIS~\cite{Koivisto2001, Tarvainen2009} operated with a single frequency (14.085~GHz) in continuous mode. 

The PE ammeter was placed on a radial diagnostics port at approximately $0.4$~m distance from the plasma. The line-of-sight plasma volume was limited by the diagnostics port ($\diameter = 5.2$~mm) and the collimator ($\diameter = 6$~mm). The line-of-sight is between the poles of the permanent magnet sextupole, i.e. the X-ray spectrum at the target is dominated by the characteristic radiation from the plasma and PE emission induced by radiation from the walls is not well observed.

The PE ammeter can be equipped with multiple targets and filters mounted to rotatable discs. Different filters can be used to limit the wavelength range of the radiation incident on the target. The cathode current is measured from the target with a picoammeter (Keithley 485). The emitted electrons are collected with an anode ring located approximately $3$~mm from the target and biased to $150$~V. The device is protected from plasma particles with an electric deflector, and the light is collimated with a $6$~mm diameter aperture. The target materials used in this study are metals which are typically found in ion sources as chamber materials (Al, Cu and stainless steel)~\cite{Kalvas2013, Kendall1986, Courteille1995}. A $1$~\textmu m thick mylar film, which is transparent for energies higher than $1$~keV, was used to study the X-ray part of the spectrum. A sapphire window, which is transparent in the wavelength range of $150$~nm -- $5$~\textmu m, was used to study the VUV-part of the spectrum.







\section{Results}
\begin{figure*}[tb]
\includegraphics[width=0.95\textwidth]{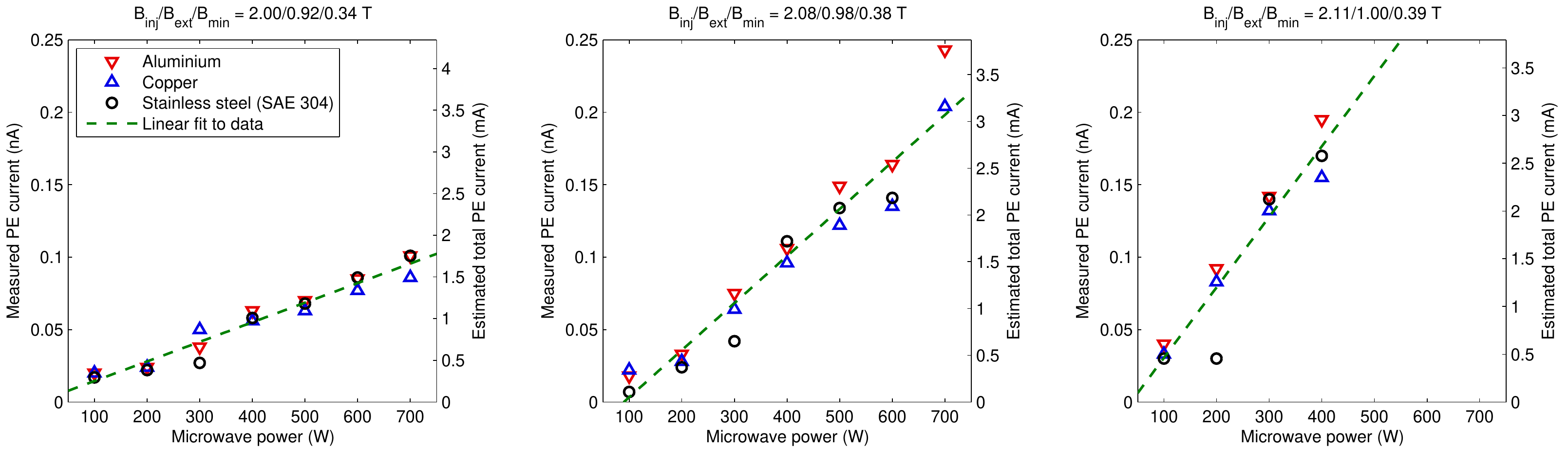}
\caption{Photoelectron current from different materials as a function of microwave power measured with residual gas at $1 \cdot 10^{-6}$~mbar pressure (injection). \label{fig:materials}}
\end{figure*}

The PE currents measured from Al, Cu and stainless steel without filters are presented in figure~\ref{fig:materials}. The PE currents are measured as a function of microwave power with different magnetic field strengths using residual gas at $1 \cdot 10^{-6}$~mbar pressure. The PE current increases linearly with increasing microwave power in the given power range. There are no significant differences between different metals.


The total PE emission from plasma chamber walls is estimated using a Monte Carlo code, which tracks single electron trajectories in the ion source magnetic field~\cite{Kalvas2014}. The particles are launched isotropically from random locations with $B < B_{\mathrm{ECR}}$, where $B_{\mathrm{ECR}}$ is the resonance field for $10$~keV electrons. Electrons are tracked $1$~\textmu s or until they hit the wall and the plasma density profile is produced from the electron population. Photons are emitted isotropically with an intensity directly proportional to the plasma density, and the probability for a single photon to hit the target is calculated. The total PE current is estimated by comparing the number of photons hitting the target to the total photon flux from the plasma. The estimated total PE current yields the maximum value from the total area of the plasma chamber wall. However, the PE flux from the wall to the plasma is limited by the magnetic field, since the cross field diffusion of the emitted electrons in transverse magnetic field is significantly slower than their propagation along the field lines. Due to the assumptions made in the calculations the given total PE currents are an order of magnitude estimates.



Different optical filters were used to find out the predominant wavelength range contributing to the measured PE emission. Filtering the radiation emitted by the residual gas plasma with the $1$~\textmu m mylar filter dropped the PE current signal to zero. Thus, it is concluded that the intensity of the X-ray emission from the plasma is too weak to cause any significant PE emission. However, the K$_{\alpha}$ for C ($277$~eV), N ($392.4$~eV) and O ($524.9$~eV) are filtered out with the $1$~\textmu m mylar filter. Similarly, when a sapphire window was placed between the plasma and the target, the PE current was too low to be measured. Hence, it is concluded that PE emission is predominantly caused by radiation emitted with energies between $8$~eV and $1$~keV.


\begin{figure*}[tb]
\includegraphics[width=0.95\textwidth]{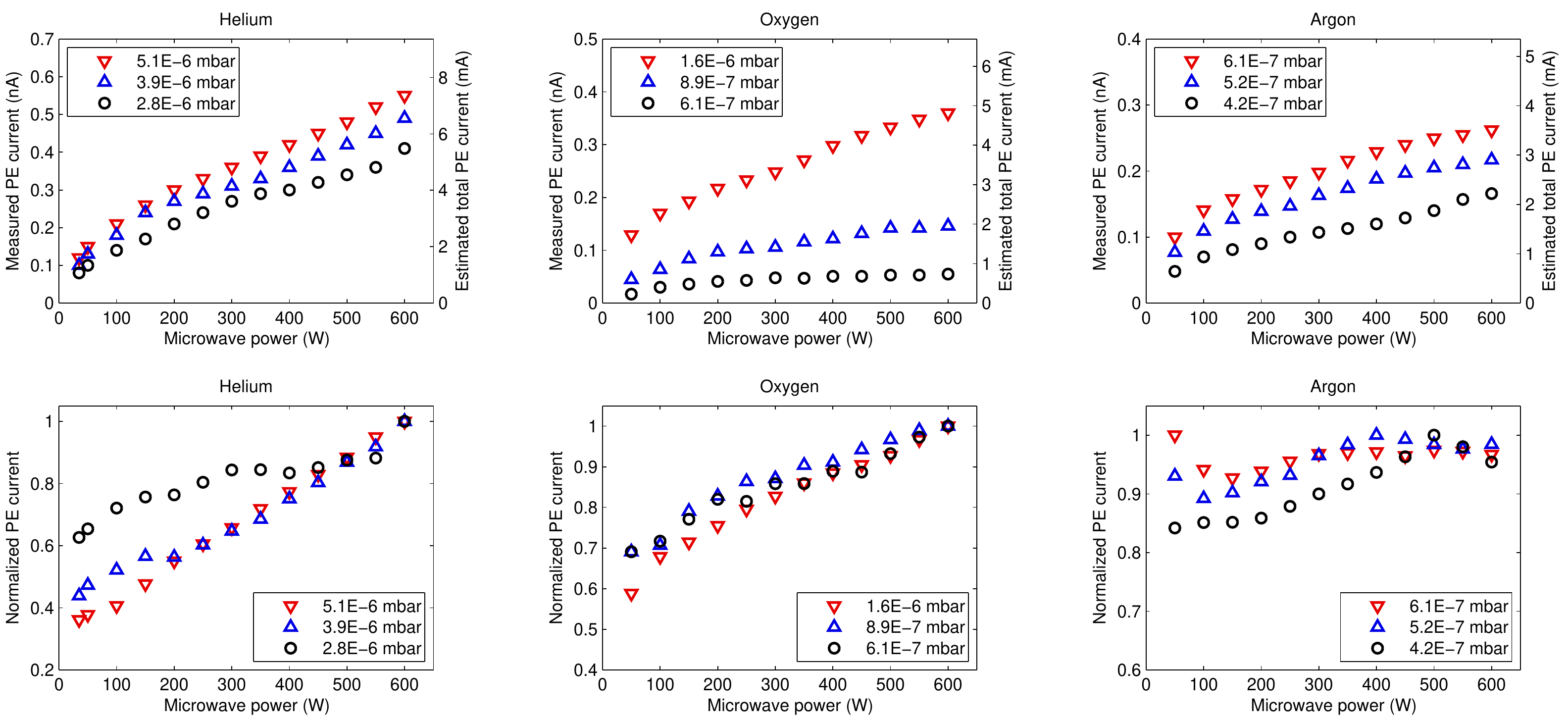}
\caption{Photoelectron current from aluminium with different gases measured with magnetic field values $B_{\mathrm{inj}}$/$B_{\mathrm{ext}}$/$B_{\mathrm{min}}$ $=$ $2.00$/$0.92$/$0.34$~T. The upper pictures show the absolute values and estimated total currents and, in the lower pictures, the measured photoelectron current is normalized with the biased disc current.\label{fig:gases}}
\end{figure*}


Higher magnetic field (coil current) corresponds to higher PE emission. The PE currents from Al, Cu and stainless steel were measured with different values of the magnetic field at the injection $B_{\mathrm{inj}}$, extraction $B_{\mathrm{ext}}$, and minimum-$B$ $B_{\mathrm{min}}$. Linear fits to measured data in figure~\ref{fig:materials} correspond to average total PE currents of $2$~mA per kW of injected microwave power, $5$~mA per kW and $7$~mA per kW for $2.00$/$0.92$/$0.34$~T, $2.08$/$0.98$/$0.38$~T and $2.11$/$1.00$/$0.39$~T magnetic fields, respectively. The increase of the PE emission with increasing magnetic field values can be caused by increased plasma density or by increased excitation rate to electronic states emitting at $8$~eV -- $1$~keV energies due to a shift of the electron energy distribution.




The PE current from Al was measured using different gases and neutral gas pressures. The results with He, O$_{2}$ and Ar plasma are presented in figure~\ref{fig:gases}. The indicated neutral particle pressures are gas calibrated readings of an ionization gauge located outside the plasma chamber and connected to it through a radial diagnostics port. The given pressure values are measured without igniting the plasma. The results indicate that the pressure dependence is significant. Measurements were performed also with Ne, Kr and Xe showing similar results.

The measured PE currents have also been normalized with the biased disc current, which can be considered as an indicator of the plasma density at a constant magnetic field. Biased disc voltage was $-120$~V. The normalized PE currents are not constant as a function of microwave power, which suggests that not only the electron density but also the excitation rate changes as a function of microwave power, which is possibly due to a shift of the electron energy distribution.






\section{Discussion}


The same Monte Carlo code that was used to estimate the total PE current has been used to estimate the total electron flux that exit the plasma~\cite{Kalvas2014}. According to the simulation $5.1$~\% of electrons with $10$~keV kinetic energy exit the plasma through the extraction aperture. The extracted ion current is typically on the order of $1$~mA. Due to ambipolar diffusion the electron current towards the extraction aperture has to be the same. Thus, the simulated total electron current from the plasma is on the order of $20$~mA, which is roughly an order of magnitude higher than the estimated maximum PE current in typical operating pressure.


Electrons are emitted from all metal surfaces inside the plasma chamber, but due to the magnetic field structure of the ECRIS only part of the emitted electrons are transmitted to the plasma. Electrons are most probably transmitted to the plasma from the area where plasma can touch the walls, which covers approximately $5$~\% of the total plasma chamber area. Hence, the total PE currents are overestimated by a factor of 20. On the other hand, the measurements in this study also underestimate the PE emission, since the contribution of direct wall bremsstrahlung and radiation emitted due to recombination of highly charged ions near the plasma chamber walls was not completely taken into account. It is probable that PE emission is not an important process in relation to the electron density of ECR plasmas, but the PE flux should be compared to the total secondary electron flux in order to conclude it's importance.

\begin{acknowledgments}
This work has been supported by the EU 7\textsuperscript{th} framework programme ``Integrating Activities -- Transnational Access'', project number: 262010 (ENSAR) and by the Academy of Finland under the Finnish Centre of Excellence Programme 2012--2017 (Nuclear and Accelerator Based Physics Research at JYFL).
\end{acknowledgments}

\bibliography{icis2015_ref}

\end{document}